\documentclass[final,5p,times,twocolumn]{elsarticle}

\usepackage{amssymb}
\usepackage{amsmath}
\usepackage{enumitem}
\usepackage{subfigure}
\usepackage{hyperref}

\hypersetup{
    colorlinks=true,
    citecolor=black,
    filecolor=black,
    linkcolor=black,
    urlcolor=black,
    pdfauthor=author,
    }

\journal{Radiation Physics and Chemistry}

\begin{document}

\begin{frontmatter}
  \title{Analysis of neutron time-of-flight spectra with a Bayesian unfolding methodology}

  \cortext[cor]{Corresponding author at CIEMAT, Avenida Complutense 40, 28040 Madrid, Spain. Tel: +34 913466615.}
  \author[label]{A.~Pérez de Rada Fiol\corref{cor}}
  \ead{alberto.rada@ciemat.es}

  \author[label]{D.~Cano-Ott}
  \author[label]{T.~Martínez}
  \author[label]{V.~Alcayne}
  \author[label]{E.~Mendoza}
  \author[label]{J.~Plaza}
  \author[label]{A.~Sanchez-Caballero}
  \author[label]{D.~Villamarín}

  \address[label]{Centro de Investigaciones Energéticas, Medioambientales y Tecnológicas (CIEMAT), 28040 Madrid, Spain}

  \begin{abstract}
    We have developed an innovative methodology for obtaining the neutron energy distribution from a time-of-flight (TOF) measurement based on the iterative Bayesian unfolding method and accurate Monte Carlo simulations. This methodology has been validated through the analysis of a realistic virtual \( \beta \)-decay experiment, including the most relevant systematic effects in a real experiment. The proposed methodology allowed for obtaining accurate results over the energy range above the neutron detection threshold.
  \end{abstract}


\end{frontmatter}

\section{Introduction}\label{intro}

Neutron spectroscopy contributes to a better understanding of nuclear physics and is an important tool in several fields like nuclear structure, nuclear astrophysics, nuclear technology, radiotherapy, radiation protection, nuclear safeguards, and fusion plasma diagnostics, among others. In the field of nuclear technology in particular, an accurate determination of neutron energy spectra is important for a variety of applications, including reactor core characterization~\cite{Fiorina2012}, reactor shielding~\cite{Hajji2021}, and nuclear waste transmutation~\cite{Abanades2002}. There are several neutron spectroscopy techniques, which can be classified by the principle used to measure the energy of the neutrons~\cite{Brooks2002}. In this paper, we will focus on the time-of-flight (TOF) technique.

The TOF technique is the main technique applied in high-resolution neutron spectroscopy. It consists of the determination of the neutron kinetic energy from the time it takes a neutron to travel the known distance---that is, the flight path---between the point where it is produced and the detection system. The neutron energy resolution (\( \Delta E / E \)) obtained with this technique depends on both the detection system's intrinsic time resolution and the resolution in the flight path~\cite{Kornilov2009}:

\begin{equation}
  \frac{\Delta E}{E} = \gamma (\gamma + 1) \sqrt{{\left( \frac{\Delta t}{t} \right)}^2 + {\left( \frac{\Delta d}{d} \right)}^2},
  \label{EnergyResolution}
\end{equation}

where:

\begin{itemize}
  \item \( \gamma \) is the Lorentz factor;
  \item \( t \) is the neutron TOF and \( \Delta t \) its uncertainty, the time resolution;
  \item \( d \) is the neutron flight path and \( \Delta d \) its uncertainty, typically approximated by half of the detector's active volume depth.
\end{itemize}

From Equation~\ref{EnergyResolution} it can be observed that the thinner and further away from the emission point the detectors are, the better the energy resolution is. However, having thin detectors covering a small solid angle decreases the neutron detection efficiency. Even though the detection efficiency can be increased by using a larger number of detectors, a compromise between the energy resolution and the detection efficiency is always necessary. The MOdular Neutron time-of-flight SpectromeTER (MONSTER)~\cite{Garcia2012, Martinez2014, Martinez2024} has been designed and built having in mind all these considerations. It is based on the BC501A and EJ301 liquid scintillators and is optimized for \( \beta \)-delayed neutron spectroscopy and other applications like \( (\alpha, n) \) and neutron-induced reactions.

MONSTER was conceived for the measurements of the energy spectrum of \( \beta \)-delayed neutrons in the range between \( 300 \) keV and \( 20 \) MeV with the TOF technique and their partial branching ratios to the different excited states in the final nucleus by applying \( \beta\textrm{-}n\textrm{-}\gamma \) coincidences. MONSTER is the result of an international collaboration between Centro de Investigaciones Energéticas, MedioAmbientales y Tecnológicas (CIEMAT), Accelerator Laboratory of the University of Jyväskylä (JYFL-ACCLAB), Variable Energy Cyclotron Center (VECC), Instituto de FÍsica Corpuscular (IFIC), and Universitat Politècnica de Catalunya (UPC) within the framework of the DEcay SPECtroscopy (DESPEC)~\cite{Rubio2006} collaboration of the NUclear STucture, Astrophysics, and Reactions (NUSTAR)~\cite{Nilsson2015} facility at Facility for Antiproton and Ion Research (FAIR)~\cite{Spiller2006, Mistry2022}. The main characteristics of MONSTER are:

\begin{itemize}
  \item a low detection threshold (\( \sim 30 \) keVee);
  \item a high intrinsic neutron detection efficiency (\( \sim 30 \) \%);
  \item discrimination between detected \( \gamma \)-rays and neutrons by their pulse shape;
  \item a good time resolution (\( \sim 1 \) ns).
\end{itemize}

The TOF technique requires both start and stop signals for the neutrons' TOF determination. The stop signal is given by MONSTER when a neutron is detected. In the case of a \( \beta \)-decay experiment, the start signal is provided by detecting the \( \beta \)-particle in coincidence with the neutron with a \( \beta \)-detector. In other kinds of experiments, like \( (\alpha, n) \) reaction measurements, for instance, the start signal would be provided by the accelerator's radio frequency.

Using the TOF technique to obtain a neutron energy distribution requires a method to solve the inverse problem given by:

\begin{equation}
  \boldsymbol{TOF} = \boldsymbol{R} \cdot \boldsymbol{E},
  \label{Inverse}
\end{equation}

where here \( \boldsymbol{TOF} \) is the TOF spectrum, \( \boldsymbol{E} \) is the neutron energy distribution, and \( \boldsymbol{R} \) is the response matrix of the detection system.

In previous works, the neutron energy spectra have been obtained by fitting the measured TOF spectra with a discrete number of time responses determined either empirically or from dedicated measurements~\cite{Perrot2006, Madurga2016, Xu2023}. In this way, the probabilities of emitting neutrons with discrete energies were obtained from the weights resulting from the fit of the corresponding time responses. This method involves various difficulties and problems such as the difficulty of obtaining, in the same conditions as the experiment, accurate time responses for a continuum of neutron energies; the appearance of numerical problems during the fit when using numerous response functions; or finding a solution that best reproduces the data in a statistical sense. In this paper, we present an innovative methodology to solve the inverse problem given by Equation~\ref{Inverse} and obtain the neutron energy spectrum from a TOF measurement. The method shares similarities with the one developed for the analysis of \( \beta \)-decay total absorption \( \gamma \)-ray spectra~\cite{Tain2007}, but relies on (neutron) response functions in the time domain, determined through precise and experimentally validated Monte Carlo simulations, and applies statistical inference to obtain the neutron energy distribution.

Determining the response function of the detector through Monte Carlo simulations allows to:

\begin{itemize}
  \item cover the whole neutron energy range of interest, maximizing the energy resolution of the result;
  \item compute the time responses for every neutron energy, instead of using a few individual responses that can be measured experimentally in any given facility or an empirical parametrization of the response function;
  \item take into account several effects such as light production of the scintillators, crosstalk between detectors, neutron scattering in surrounding materials and ancillary detectors, and others;
  \item study the impact of several kinds of systematic effects such as detection thresholds, background contributions, systematic uncertainties, and others on the solution.
\end{itemize}

It is worth noting that although the methodology presented in this paper has been developed and validated to obtain \( \beta \)-delayed neutron energy distributions with MONSTER, its extension to other TOF spectrometers and physics problems should be possible.

\begin{figure}[tb]
  \centering
  \subfigure{\includegraphics[width=\linewidth]{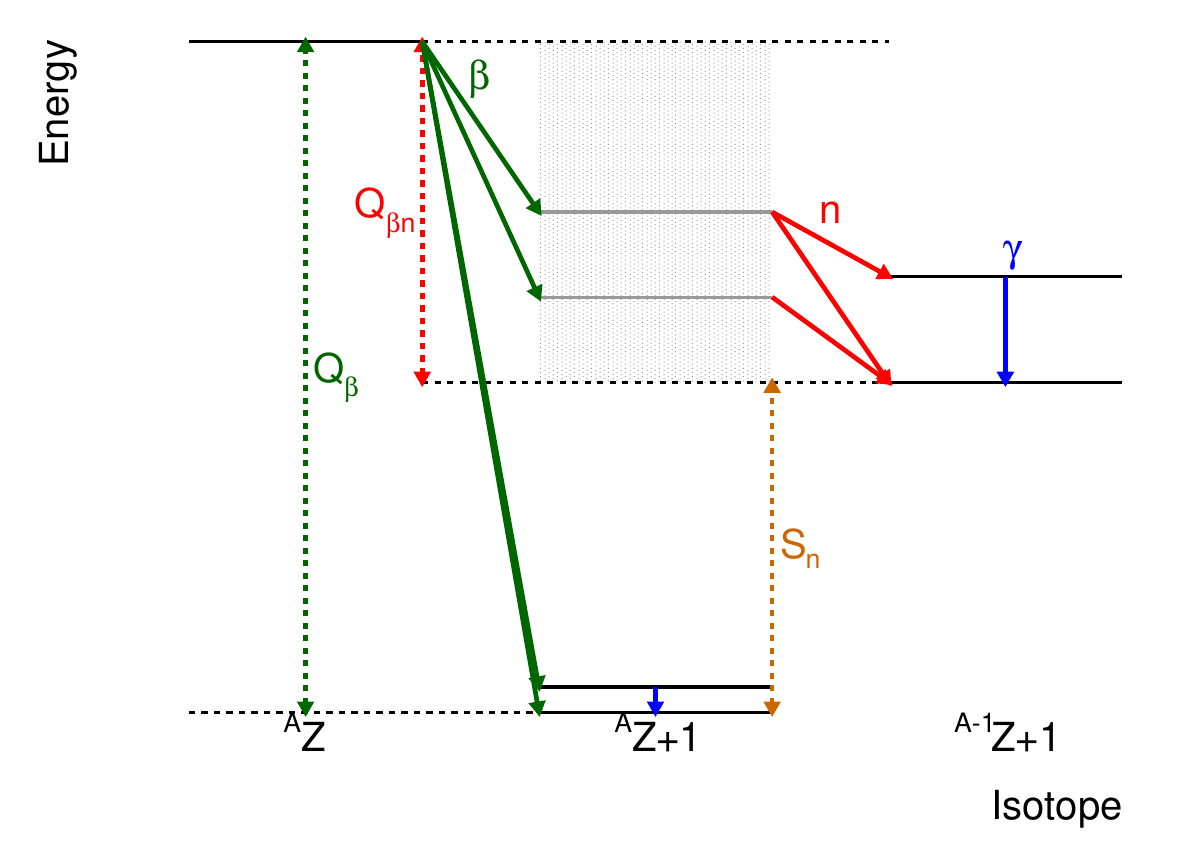}}
  \caption{\( \beta \)-delayed neutron emission scheme.}\label{Decay}
\end{figure}

\subsection{\texorpdfstring{\( \beta \)}{Beta}-delayed neutron emission}\label{decay}

\( \beta \)-delayed neutron emission phenomena occur in the neutron-rich side of the chart of nuclides. In nuclei far from the valley of stability, the available energy for a \( \beta \)-decay (\( Q_{\beta} \)) increases and the neutron separation energy (\( S_n \)) decreases, making this decay mode possible when states over the neutron separation energy of the daughter nucleus are fed by the \( \beta \)-decay. At that point, the decay product can emit neutrons or undergo regular \( \gamma \)-deexcitation. In addition, the emitted neutrons can populate either the ground or excited states in the granddaughter nucleus. In the latter case, even further emission of neutrons or regular \( \gamma \)-deexcitation would follow. The \( \beta \)-delayed neutron emission process is schematized in Figure~\ref{Decay}.

In this paper, the experimental setup modeled for presenting and validating the analysis methodology corresponds to the one used in the measurement of the \( \beta \)-delayed neutron spectra of \( ^{85, 86} \)As at JYFL-ACCLAB with MONSTER in March 2019~\cite{ND2022, EuNPC2022, PRC}. The experimental results will be presented in a forthcoming paper.

\subsection{TOF spectrum unfolding}\label{tof}

We have developed a neutron TOF spectrum analysis methodology based on the iterative Bayesian unfolding~\cite{DAgostini1995} of the experimental data with a response matrix calculated with very accurate Monte Carlo simulations. Based on Bayes' theorem, this method allows for the statistical inference of a true distribution (neutron energy distribution) from a folded one (TOF spectrum) given a known distortion of the true physical quantity (response matrix). As a result, it produces the neutron energy distribution and the total number of neutrons emitted above the neutron detection threshold.

Equation~\ref{Inverse} can be expressed in terms of \( n_C \) independent causes (\( C_i \), the \( \beta \)-delayed neutron energy distribution), \( n_E \) possible effects (\( E_j \), the TOF spectrum), and the conditional probability that a given cause produces a certain effect (\( P(E_j|C_i) \), the response matrix). By doing so, the inverse problem can be expressed with the ingredients of the Bayes theorem:

\begin{equation}
  P(C_i|E_j) = \frac{P(E_j|C_i)P_0(C_i)}{\sum_{l=1}^{n_C} P(E_j|C_l)P_0(C_l)},
  \label{BayesNEffects}
\end{equation}

where:

\begin{itemize}
  \item \( P(C_i|E_j) \) is the conditional probability that a given effect has been produced by a certain cause;
  \item \( P_0(C_i) \) is the initial probability of the causes;
  \item \( \sum_{i=1}^{n_C} P_0(C_i) = 1 \), which means that a non-existing cause cannot be invented;
  \item \( \sum_{i=1}^{n_C} P(C_i|E_j) = 1 \), meaning that any observed effect must come from one or more of the considered causes;
  \item \( 0 \leq \epsilon_i \equiv \sum_{j=1}^{n_E} P(E_j|C_i) \leq 1 \), with \( \epsilon_i \) being the efficiency of detecting a given cause, means that it is not required for each cause to produce at least one of the considered effects.
\end{itemize}

At first sight, solving this formula poses two problems: \( P(C_i|E_j) \) depends both on the unknown \( P_0(C_i) \) and \( P(E_j|C_i) \). To solve the first one, a uniform distribution can be assumed as the starting point, and an iterative process can be used to increase the knowledge of \( P(C_i) \) with every iteration. As for the second one, \( P(E_j|C_i) \) can be estimated with Monte Carlo simulations. Lastly, the effect of systematic uncertainties on the results can be evaluated by trying different estimations of the response matrix.

If \( n(E_j) \) events are observed with effect \( E_j \), and taking into account the inefficiency---that is, if \( \epsilon_i = 0 \), which means that the cause \( C_i \) cannot be detected, \( \hat{n}(C_i) \) will be set to zero---the expected number of events that can be assigned to each cause \( C_i \), \( \hat{n}(C_i) \), is:

\begin{equation}
  \hat{n}(C_i) = \frac{1}{\epsilon_i} \sum_{j=1}^{n_E} n(E_j)P(C_i|E_j), \quad \epsilon_i \neq 0.
  \label{TrueEventsPerCause}
\end{equation}

Finally, from Equation~\ref{TrueEventsPerCause}, the estimated true total number of events, \( \hat{N}_{true} \), is:

\begin{equation}
  \hat{N}_{true} = \sum_{i=1}^{n_C} \hat{n}(C_i),
  \label{TrueEvents}
\end{equation}

and the final probabilities of the causes, \( \hat{P}(C_i) \), can be obtained as:

\begin{equation}
  \hat{P}(C_i) \equiv P(C_i|\boldsymbol{n}(E)) = \frac{\hat{n}(C_i)}{\hat{N}_{true}},
  \label{CausesProbabilities}
\end{equation}

where \( \boldsymbol{n}(E) \equiv \{n(E_1), n(E_2), \ldots n(E_{n_E})\} \) is the distribution of frequencies after \( N_{obs} \) observations.

Starting from a uniform distribution of the initial probabilities of the causes---that is, \( P_0(C_i) = 1 / n_C \), and hence the initial number of events being \( n_0(C_i) = P_0(C_i)N_{obs} \)---, the iterative method proceeds as follows:

\begin{enumerate}
  \item calculate the expected distributions of the number of events, \( \hat{\boldsymbol{n}}(C) \), and of the probabilities of the causes, \( \hat{\boldsymbol{P}}(C) \);
  \item make a \( \chi^2 \) comparison between \( \hat{\boldsymbol{n}}(C) \) and \( \boldsymbol{n}_0(C) \), with \( \boldsymbol{n}_0(C) \) being the distribution of the initial number of events;
        \begin{enumerate}[label*=\arabic*.]
          \item if the \( \chi^2 \) comparison between \( \hat{\boldsymbol{n}}(C) \) and \( \boldsymbol{n}_0(C) \) is small enough, which means that the solution is stable, stop the process;
        \end{enumerate}
  \item replace \( \boldsymbol{P}_0(C) \) by \( \hat{\boldsymbol{P}}(C) \), \( \boldsymbol{n}_0(C) \) by \( \hat{\boldsymbol{n}}(C) \), and start again.
\end{enumerate}

This unfolding method presents several advantages over others, such as:

\begin{itemize}
  \item it is theoretically well grounded, it deals directly with distributions, and it does not require matrix inversion;
  \item the causes and the effects can refer to different domains, like neutron energy and TOF\@;
  \item different binning can be used for the causes and the effects;
  \item any kind of smearing and migration from the causes to the effects can be taken into account;
  \item starting the process with no knowledge of the distribution of the causes leads to accurate results;
  \item different background sources can be taken into account by just including them as extra causes;
  \item it provides the covariance matrix of the results.
\end{itemize}

\section{Monte Carlo simulations}\label{sims}

As mentioned in Section~\ref{tof}, the response matrix \( P(E_j|C_i) \) can be calculated through Monte Carlo simulations. In this work, the Geant4 simulation toolkit~\cite{Agostinelli2003} was used to determine the TOF response of the entire experimental setup to individual neutron energy ranges. Due to the complex nature of neutron interactions, a good physics model and correct implementation of the detector geometries and surrounding materials are very important for calculating accurate TOF distributions. For this reason, the G4ParticleHP model~\cite{Mendoza2011} in Geant4 was used, since it has been validated extensively for the simulation of MONSTER modules~\cite{Martinez2014, Garcia2011, Garcia2017}. This high-precision model includes the necessary models and data for a complete description of neutron-induced alpha production reactions on carbon.

The detailed simulations performed in this work include the parametrized light output functions (\( L \), in MeVee) of the scintillators for electrons, protons, and heavier ions as a function of their energy (\( E \), in MeV). For electrons, the light output function was taken from Reference~\cite{Novotny1997}. For protons and heavier ions, the light output functions were selected from the wide set of functions available in the data files of NRESP7.1~\cite{Dietze1982}, and adapted for the simulation of MONSTER modules by comparing the results of simulations against experimental measurements of response functions for neutrons of different energies:

\begin{equation}
  \begin{array}{ll}
    L_{p}      & = -1.4177 + 0.68639E_{p} \qquad E_{p} > 8 \textrm{ MeV} \\
    L_{d}      & = 2L_{p}(0.5E_{d})                                      \\
    L_{\alpha} & =
    \begin{cases}
      0.02076E_{\alpha}^{1.871}    & E_{\alpha} < 6.76 \textrm{ MeV}    \\
      -0.70929 + 0.21454E_{\alpha} & E_{\alpha} \geq 6.76 \textrm{ MeV}
    \end{cases}    \\
    L_{Be}     & = 0.01E_{Be}                                            \\
    L_{B,C}    & = 0.0097E_{B,C}.
  \end{array}\label{LightOutput}
\end{equation}

For protons below \( 8 \) MeV, the light yield is calculated by linear interpolation of a set of discrete experimental values~\cite{Schmidt2002}.

In the simulation, all secondary particles giving rise to light in the scintillator are tracked, and the light yield is calculated at the end of each step using the corresponding light output function. The total light yield for each event is obtained by the addition of the light produced in each step by each kind of secondary particle produced. In this way, the simulations were able to reproduce the experimental response functions within a \( 5 \) \% of uncertainty. Including the light production mechanisms of the scintillators in the simulation allows to take into account important effects such as the neutron detection threshold of the detectors due to the produced light, which in turn allows to calculate the neutron detection efficiency with high accuracy.

\begin{figure}[tbp]
  \centering
  \subfigure[Full setup]{\includegraphics[width=\linewidth,trim={3cm 2cm 3cm 2cm},clip]{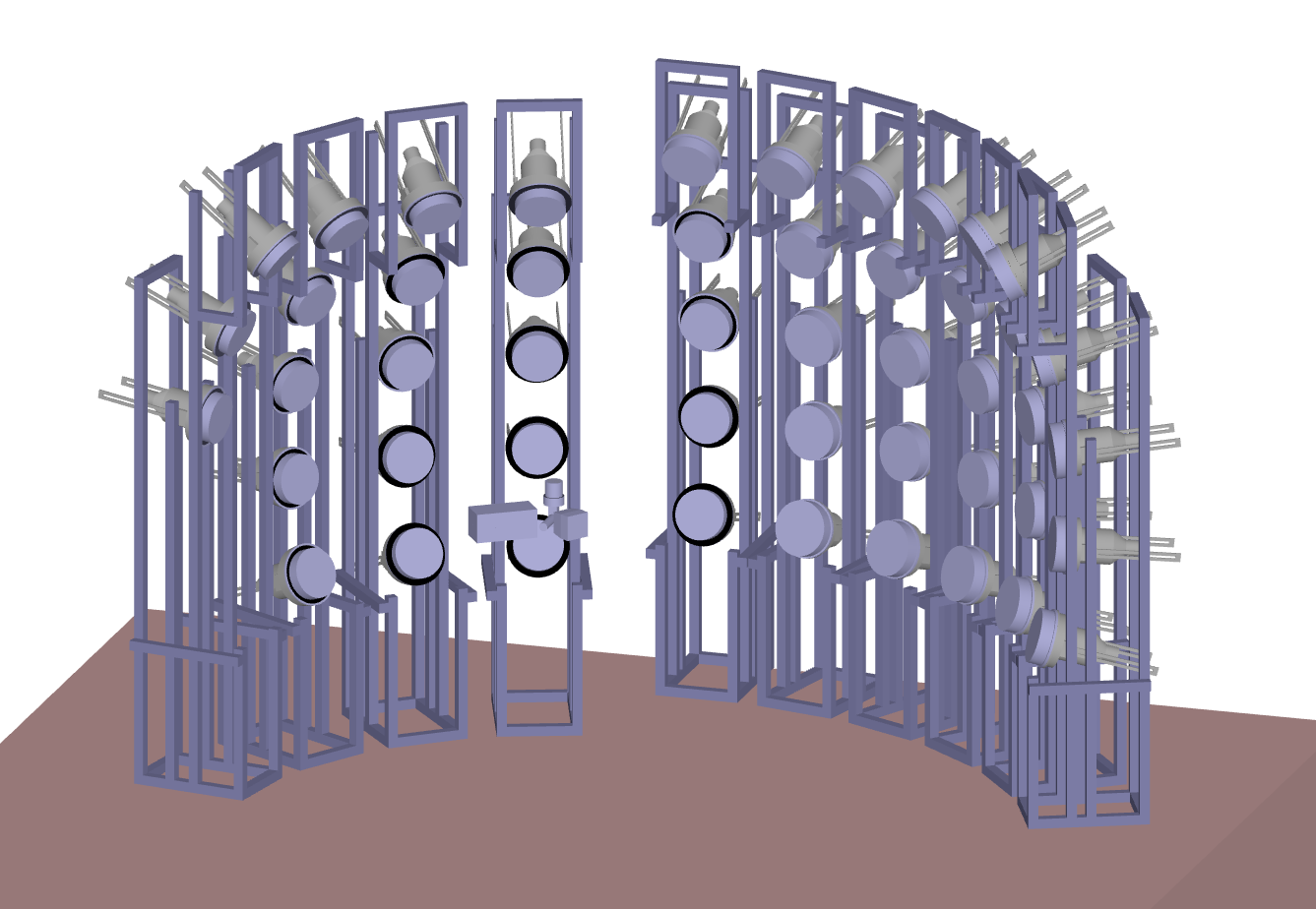}}
  \subfigure[MONSTER cell]{\includegraphics[angle=90,width=\linewidth,trim={0cm 0cm 7.25cm 0cm},clip]{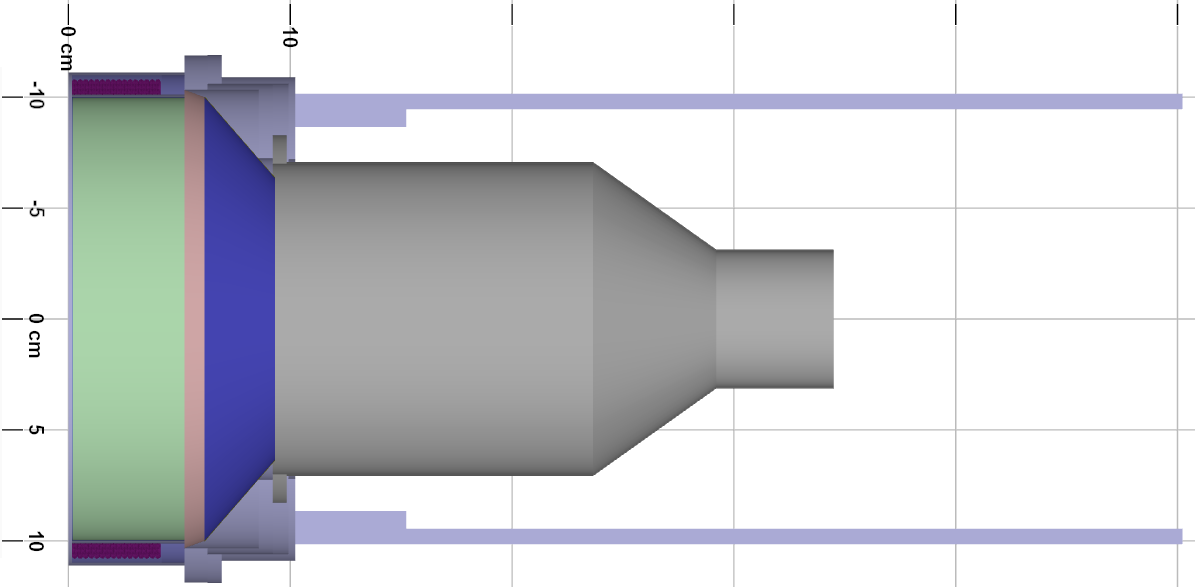}}
  \caption{Geometric model of the experimental setup simulated with Geant4. Panel (a) shows the full experimental setup, with 30 MONSTER modules at 2 m, 18 MONSTER modules at 1.5 m, and some ancillary detectors surrounding the implantation point. Panel (b) shows the detailed geometry of a MONSTER module: the aluminum cover and supports, and the mu-metal shielding in different shades of gray; the BC501A liquid scintillator in green; in salmon the quartz window; in blue the PMMA light guide; the thin layers of reflective paint, surrounding the scintillator and the light guide, in blue and yellow, respectively; and in magenta the BC501A expansion volume.}\label{SimulationSetup}
\end{figure}

About the geometric model, Figure~\ref{SimulationSetup} shows the high level of detail with which MONSTER was implemented in the simulation, with modules placed at \( 1.5 \) m and \( 2 \) m flight paths from the source. For simplicity, the analysis discussed in this paper considers only the MONSTER array at a flight path of \( 2 \) m.

\begin{figure}[tbp]
  \centering
  \subfigure{\includegraphics[width=\linewidth]{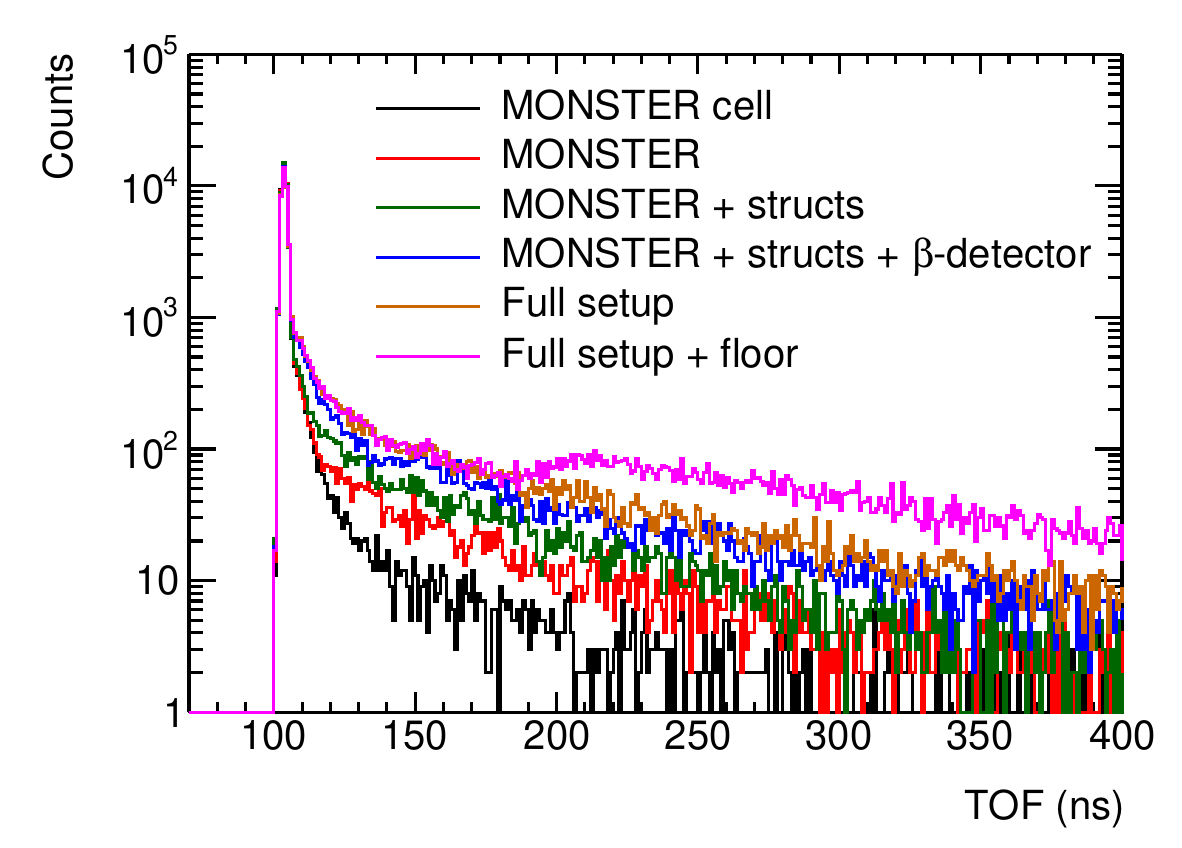}}
  \caption{MONSTER simulated TOF response to \( 2 \) MeV neutrons for several setups, including more or fewer geometry elements in the simulation.}\label{ResolutionFunction}
\end{figure}

\subsection{The Monte Carlo simulation of the TOF response function}\label{geom}

Constructing an accurate response matrix is a key step of the developed methodology. The importance of having a good geometric description in the Monte Carlo simulations is illustrated in Figure~\ref{ResolutionFunction}, which shows the TOF response (also called TOF resolution function) of MONSTER at a flight path of \( 2 \) m for \( 2 \) MeV neutrons simulated with different geometries. The TOF responses have been convoluted with a Gaussian time resolution of \( 1.4 \) ns at full width at half maximum. Each of the simulated geometries includes the elements of the previous one plus additional ones. The response of a single module (``MONSTER cell'', in black) shows a broad peak around the \( 2 \) MeV neutron TOF and a tail at higher times, due to multiple neutron interactions inside the detector (active volume and passive materials). The response of all modules (``MONSTER'', in red) shows a more pronounced tail at longer times than the single module, due to the crosstalk with the neighboring detectors. The simulation including the aluminum structures that hold the detectors (``MONSTER + structs'', in green) shows again an enhancement of the tail, due to neutron scattering in the aluminum. The effect of the \( \beta \)-detector (``MONSTER + structs + \( \beta \)-detector'', in blue) further enhances the tail at longer times. All these effects are also present in the simulation including the \( \gamma \)-ray detectors (``Full setup'', in orange). Last, but not least, the floor (``Full setup + floor'', in magenta) acts as a neutron reflector and introduces a bump in the tail due to scattered neutrons arriving at longer times than those corresponding to their energy.

Thus, it can be concluded that the geometric details of the experimental setup are relevant and that they need to be included in the Monte Carlo simulations for constructing an accurate response matrix.

\section{Analysis of a simulated experiment}\label{analsimexp}

\begin{figure}[ht]
  \centering
  \subfigure[Energy distribution]{\includegraphics[width=\linewidth]{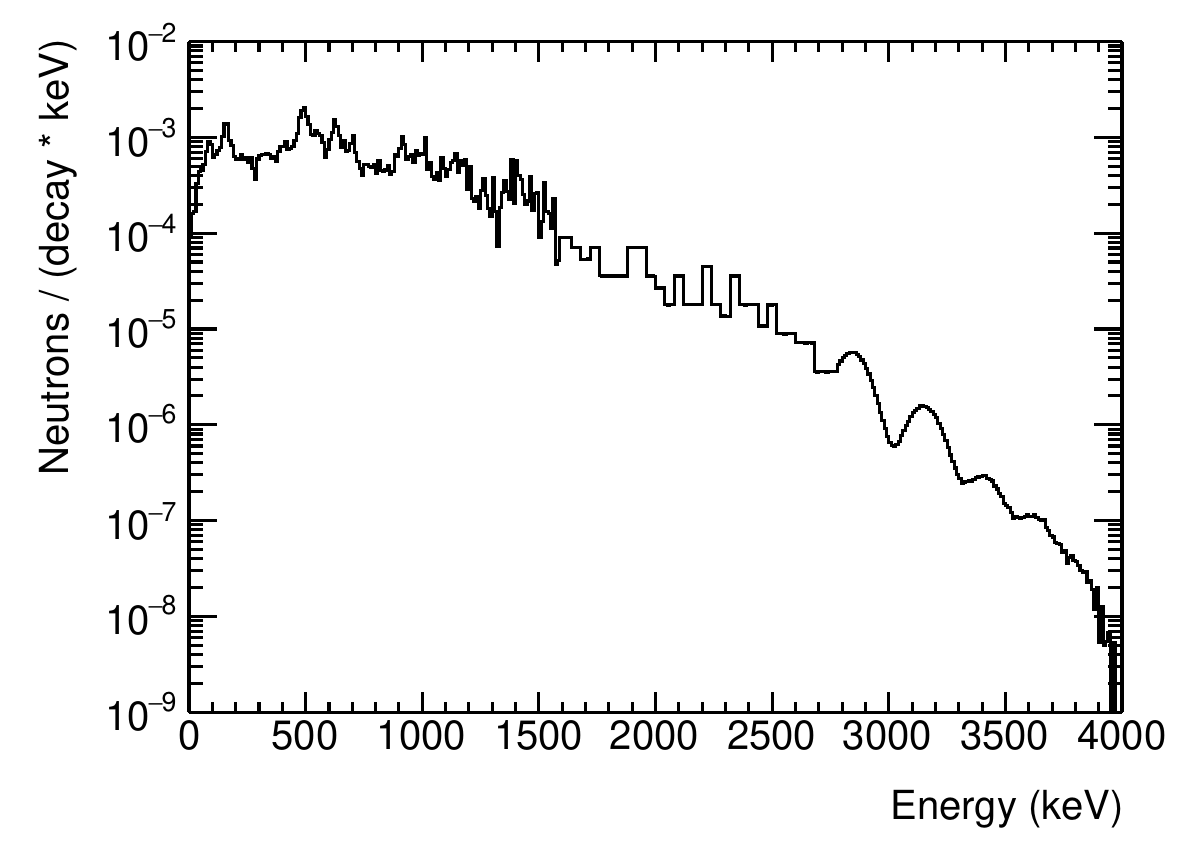}}
  \subfigure[TOF]{\includegraphics[width=\linewidth]{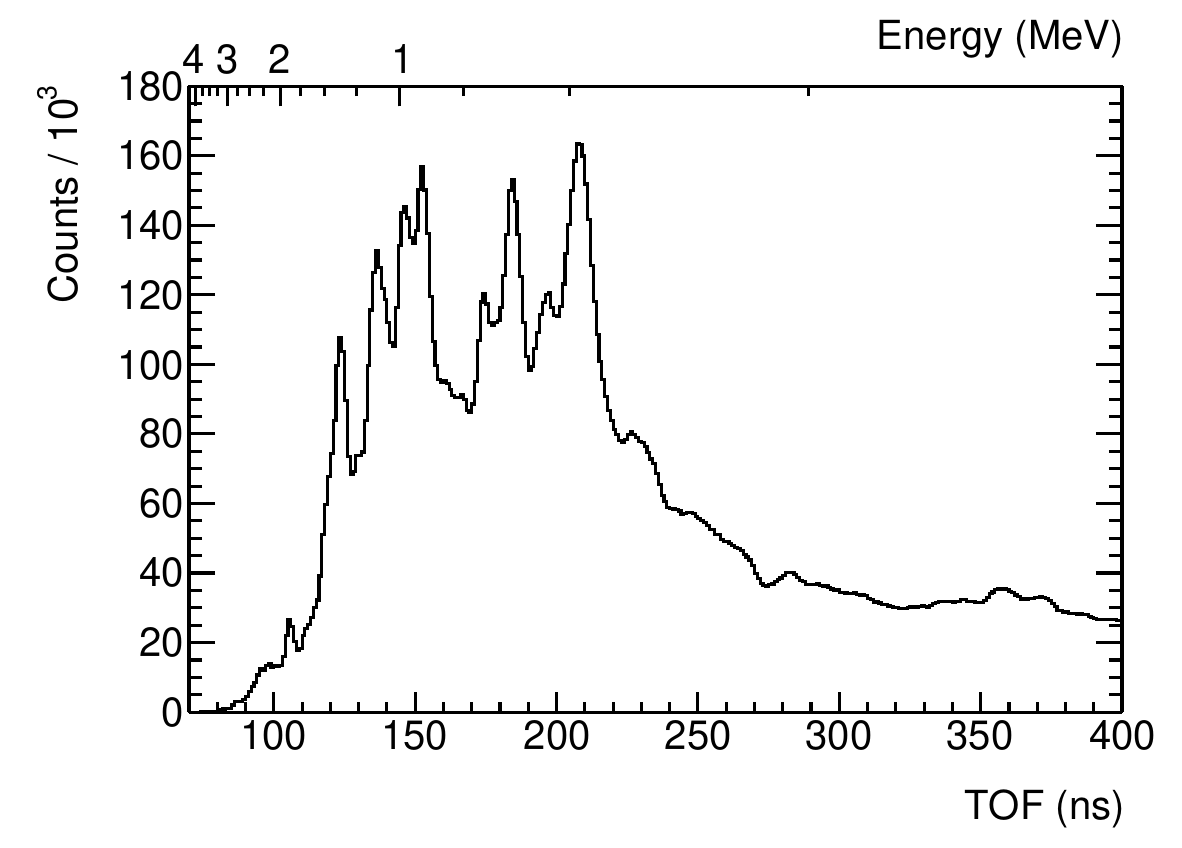}}
  \caption{Panel (a) shows the energy distribution of the neutrons emitted in the \( \beta \)-decay of \( ^{85} \)As available in ENDF/B-VIII.0 and used in the simulations. Panel (b) shows the TOF spectrum obtained from the Monte Carlo simulations.}\label{Simulation}
\end{figure}

The unfolding procedure has been validated with the analysis of a Monte Carlo simulated experiment. The detection of the neutrons emitted in the \( \beta \)-decay of \( ^{85} \)As was simulated with Geant4. For this purpose, the neutron energy distribution---that is, the true distribution of causes---shown in the top panel of Figure~\ref{Simulation} was taken from the ENDF/B-VIII.0~\cite{Brown2018, Brady1989} evaluated nuclear data library. The TOF spectrum resulting from the simulation---that is, the distribution of effects---is shown in the bottom panel of the figure. The impact on the results of the analysis due to discretization of the response matrix, the neutron detection threshold, and the \( \gamma \)-ray background present in the TOF spectrum are described in the subsections following. Last, a complete experiment including the effect of the \( \beta \)-detection efficiency is discussed.

\subsection{Impact of the binning of the response matrix}\label{bin}

The unfolding procedure requires to discretize the response matrix in TOF and incident neutron energy. For the case of the incident neutron energy, different types of binning were investigated to evaluate which one allows to infer the initial neutron energy distribution more accurately after the unfolding.

\begin{figure}[tbp]
  \centering
  \subfigure[\( \Delta E \)-binning]{\includegraphics[width=\linewidth]{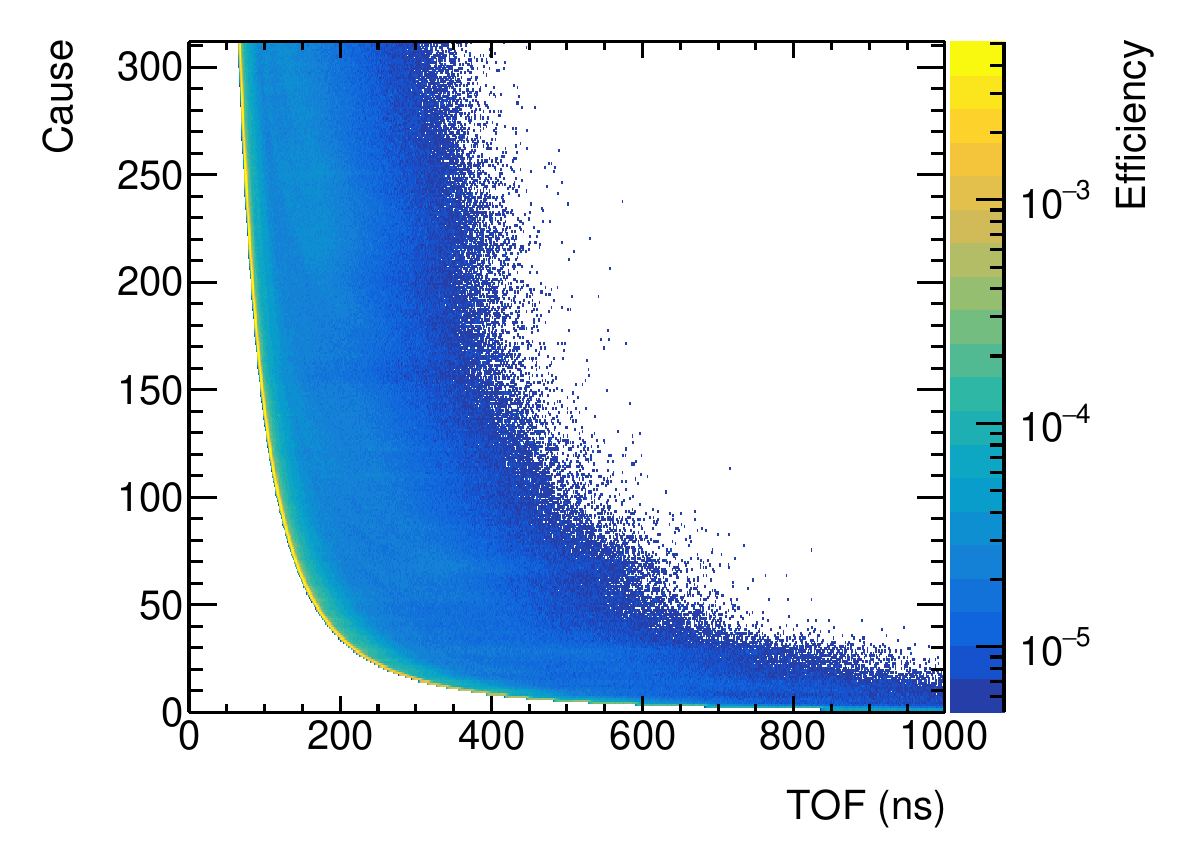}}
  \subfigure[\( \Delta t \)-binning]{\includegraphics[width=\linewidth]{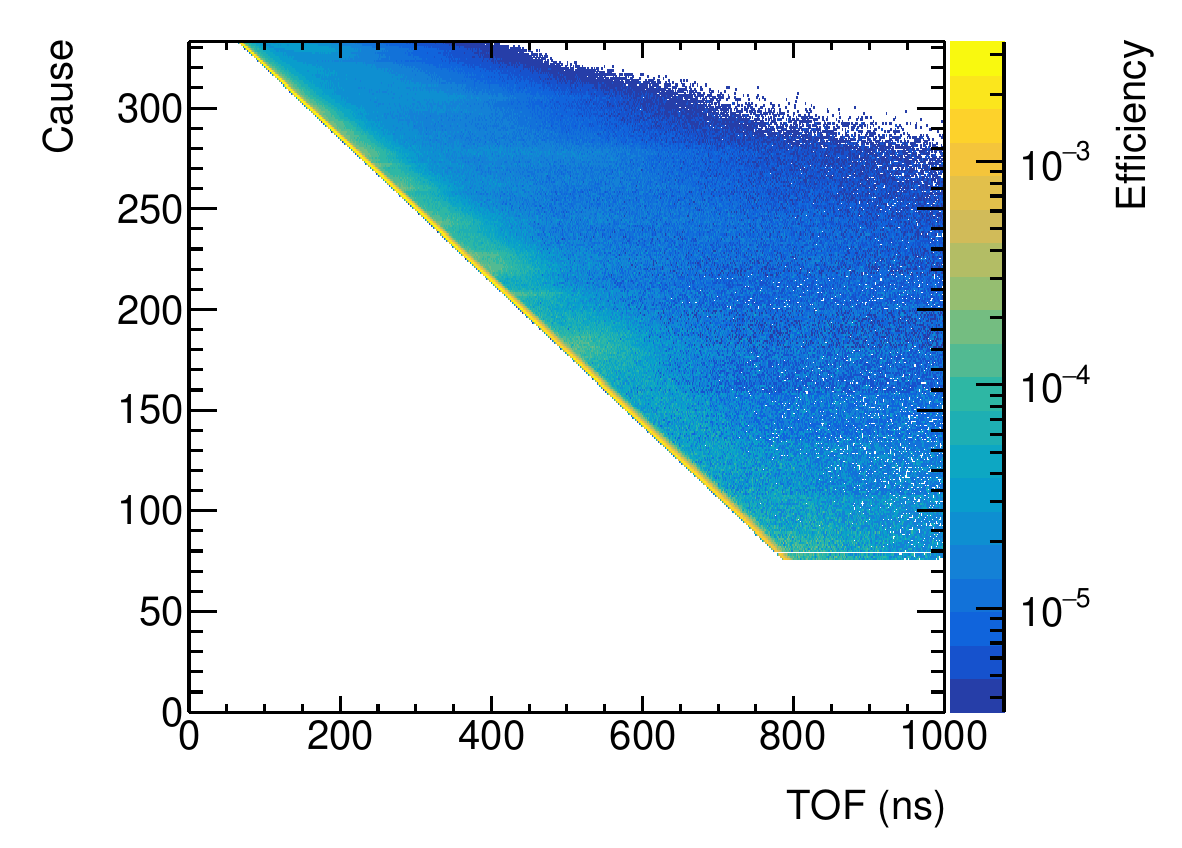}}
  \subfigure[\( \Delta R \)-binning]{\includegraphics[width=\linewidth]{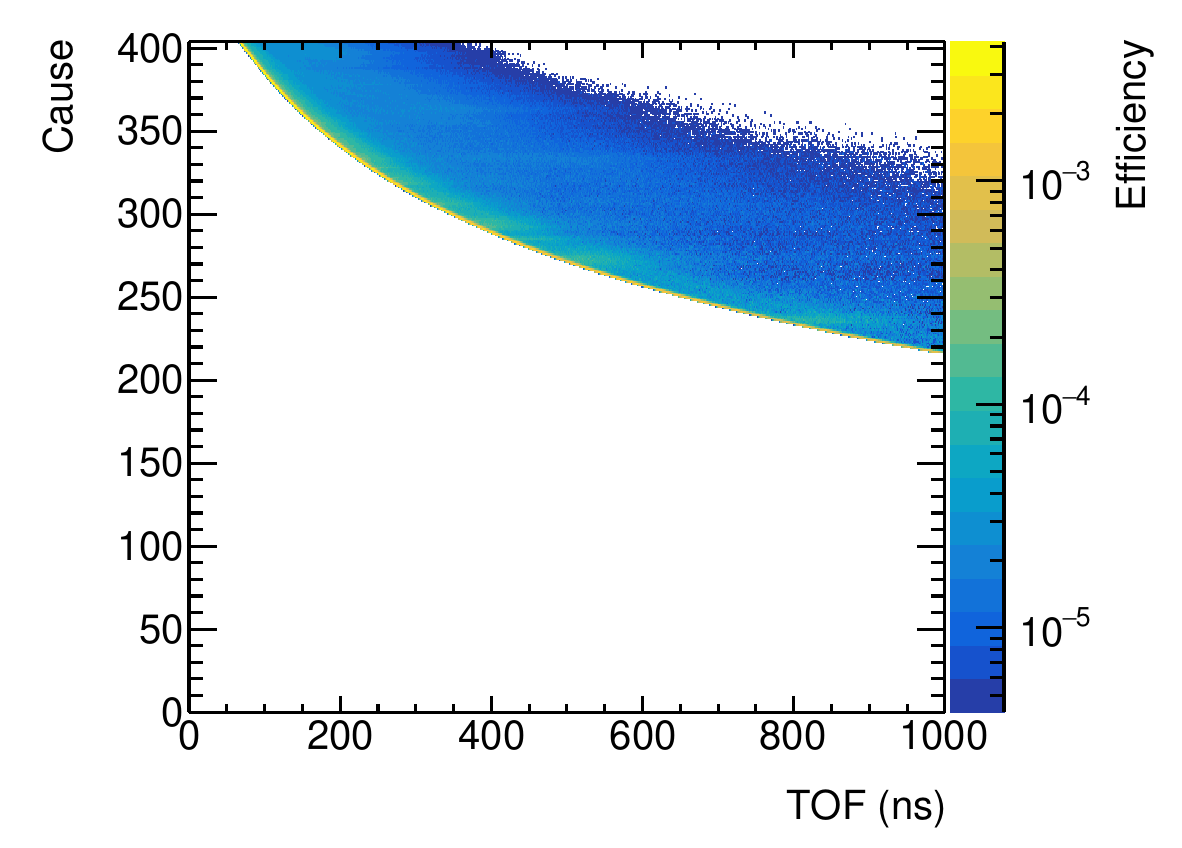}}
  \caption{Response matrices obtained with the different types of binning studied. Each cause bin corresponds to a neutron energy range of width following the respective binning type.\vspace{\baselineskip}\vspace{\baselineskip}}\label{ResponseMatrices}
\end{figure}

Figure~\ref{ResponseMatrices} shows the response matrices with three different types of binning. The vertical axis corresponds to the causes. In this case, each cause corresponds to a range of neutron energies. The horizontal axis corresponds to the effects---that is, the TOF\@. Several values of bin widths were studied for each type of binning but, for clarity, only the most relevant are presented:

\begin{itemize}
  \item \( \Delta E \)-binning: cause bins of constant width in energy of \( 15 \) keV (top panel of the figure).
  \item \( \Delta t \)-binning: cause bins with energy widths corresponding to constant widths in time of \( 2.8 \) ns (middle panel of the figure).
  \item \( \Delta R \)-binning: cause bins with energy widths that correspond to the energy resolution of the detection system (\( \Delta E / E \), given by Equation~\ref{EnergyResolution}) evaluated at the low edge of the bin (bottom panel of the figure).
\end{itemize}

As can be seen in the figure, the different types of binning produce response matrices with different structures. The narrow peaks of the response functions are observed in warmer colors, from yellow to green, while the tails of the response correspond to the light-to-darker blue regions. Panel (a), which corresponds to the \( \Delta E \)-binning, shows the non-linear inverse dependence of the neutron energy (causes) on its TOF\@. The equal spacing in energies results in numerous response functions covering a narrow TOF range and fewer response functions for the low TOF range. In the case of the \( \Delta t \)-binning, shown in panel (b) of the figure, the responses cover the whole TOF range uniformly and thus the cause-to-TOF relation is linear. Panel (c) of the figure shows the response functions in the case of the \( \Delta R \)-binning, covering the whole TOF range with a distribution following the dependence of the neutron energy resolution on its TOF\@.

\begin{figure}[tbp]
  \centering
  \subfigure[\( \Delta E \)-binning]{\includegraphics[width=\linewidth]{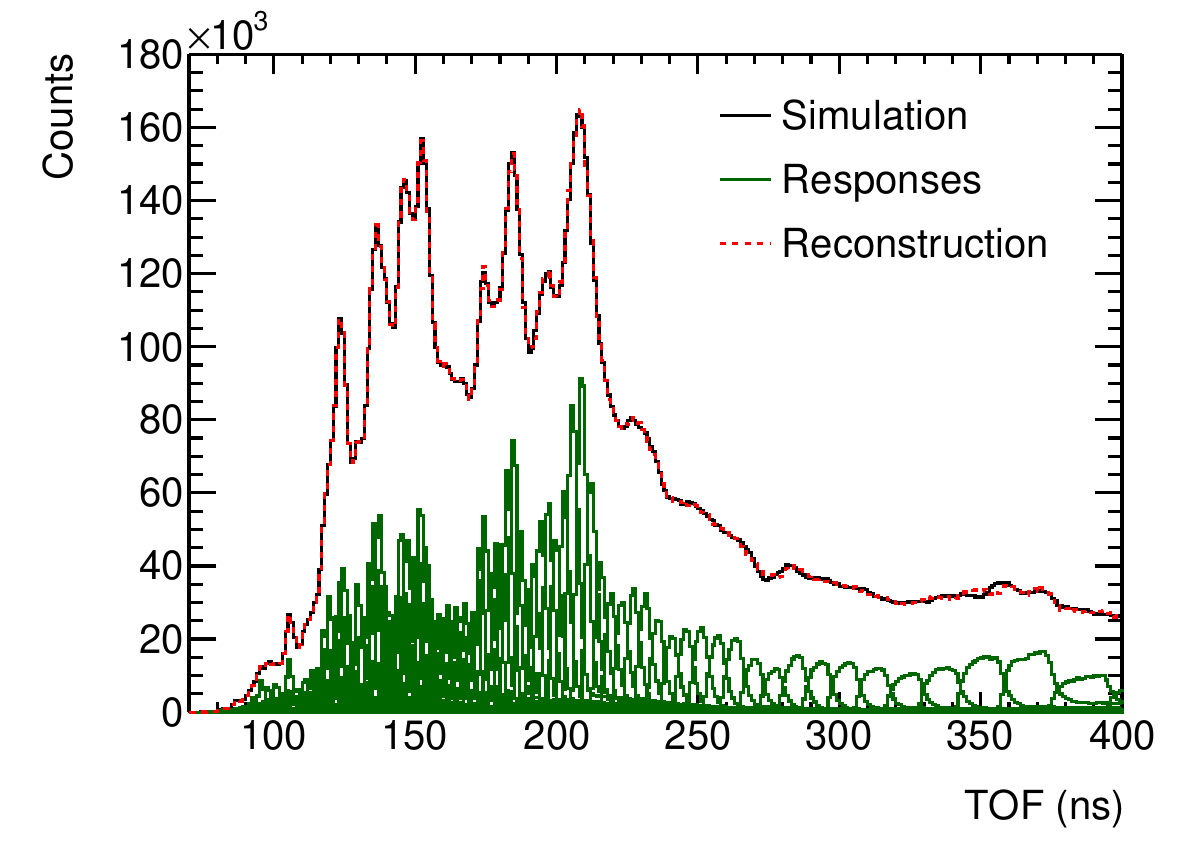}}
  \subfigure[\( \Delta t \)-binning]{\includegraphics[width=\linewidth]{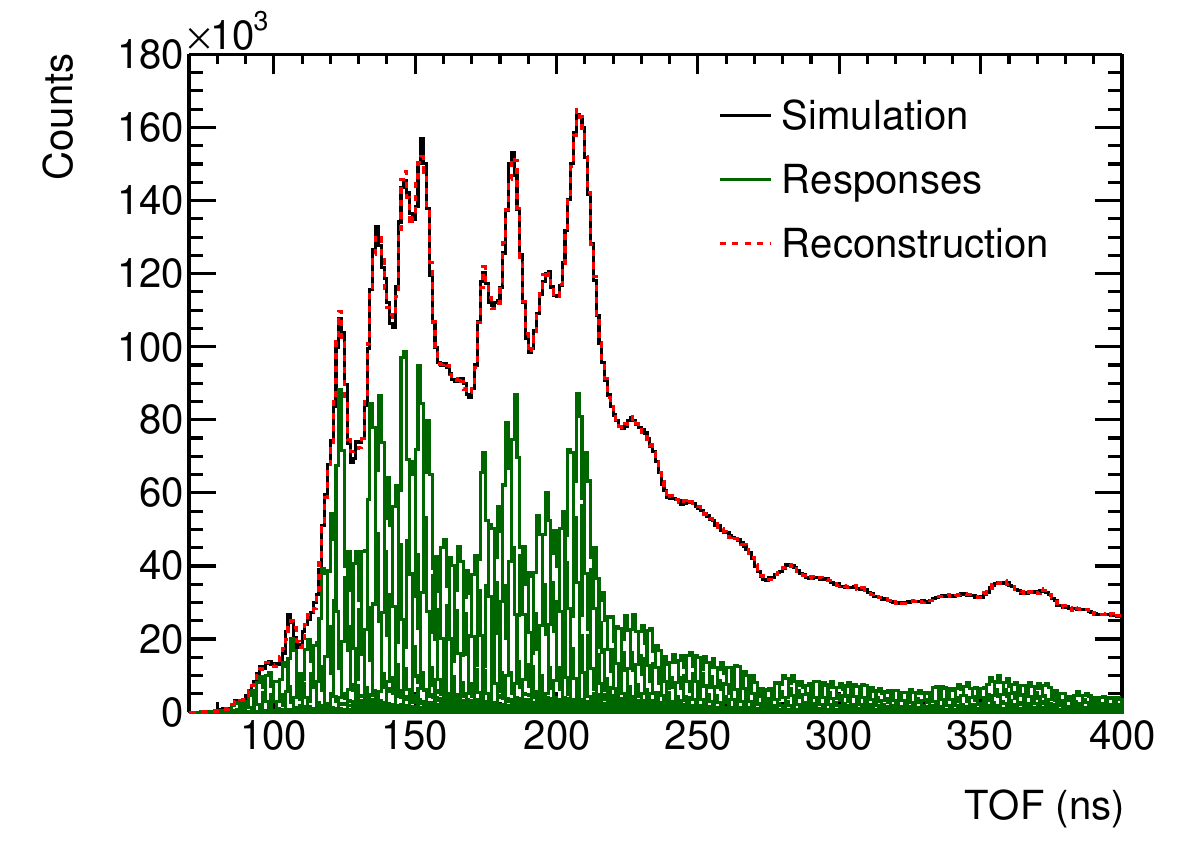}}
  \subfigure[\( \Delta R \)-binning]{\includegraphics[width=\linewidth]{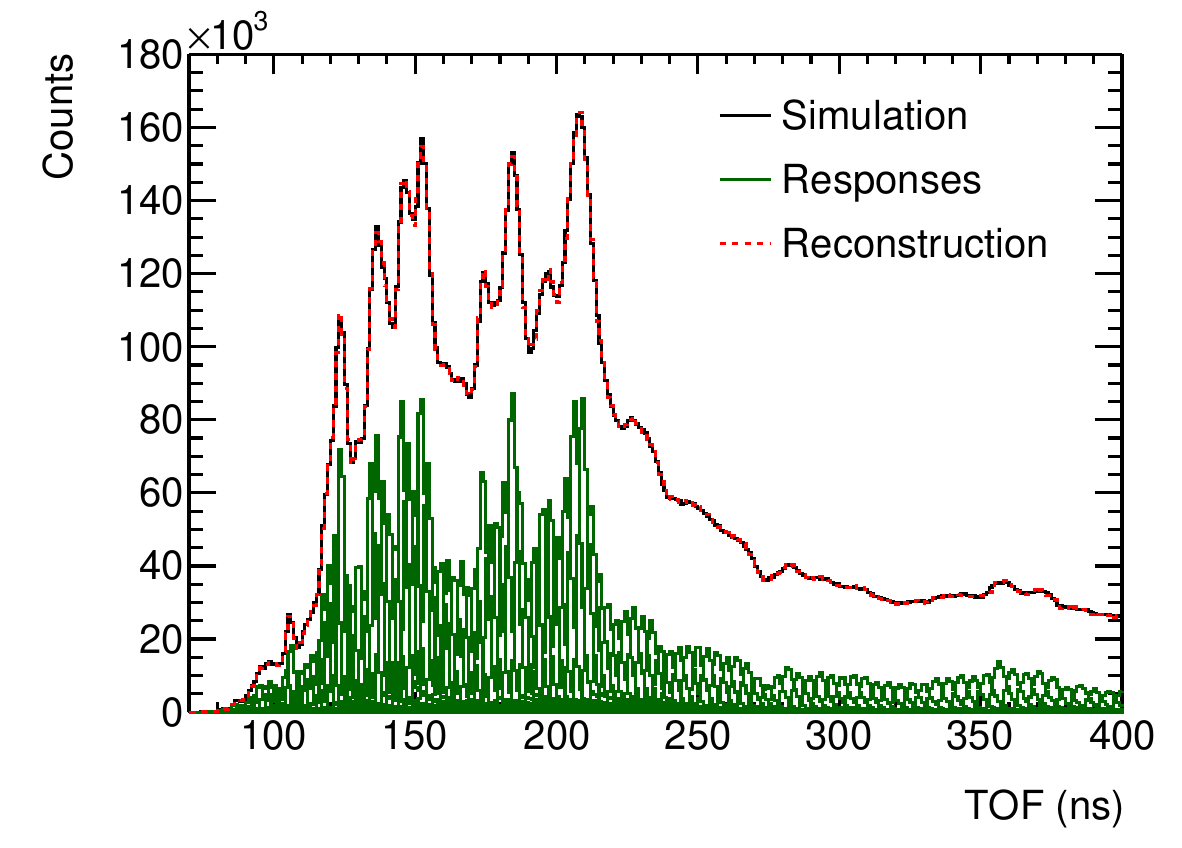}}
  \caption{TOF reconstructions obtained with the response matrices built for the different types of binning studied. The simulated TOF spectrum (``Simulation'') is shown in black, the reconstructed TOF spectrum (``Reconstruction'') in dashed red, and the response functions for the neutron energy ranges corresponding to the causes (``Responses'') in green.}\label{Reconstructions}
\end{figure}

\begin{figure*}[tbp]
  \centering
  \subfigure[\( \Delta E \)-binning]{\includegraphics[width=0.49\linewidth]{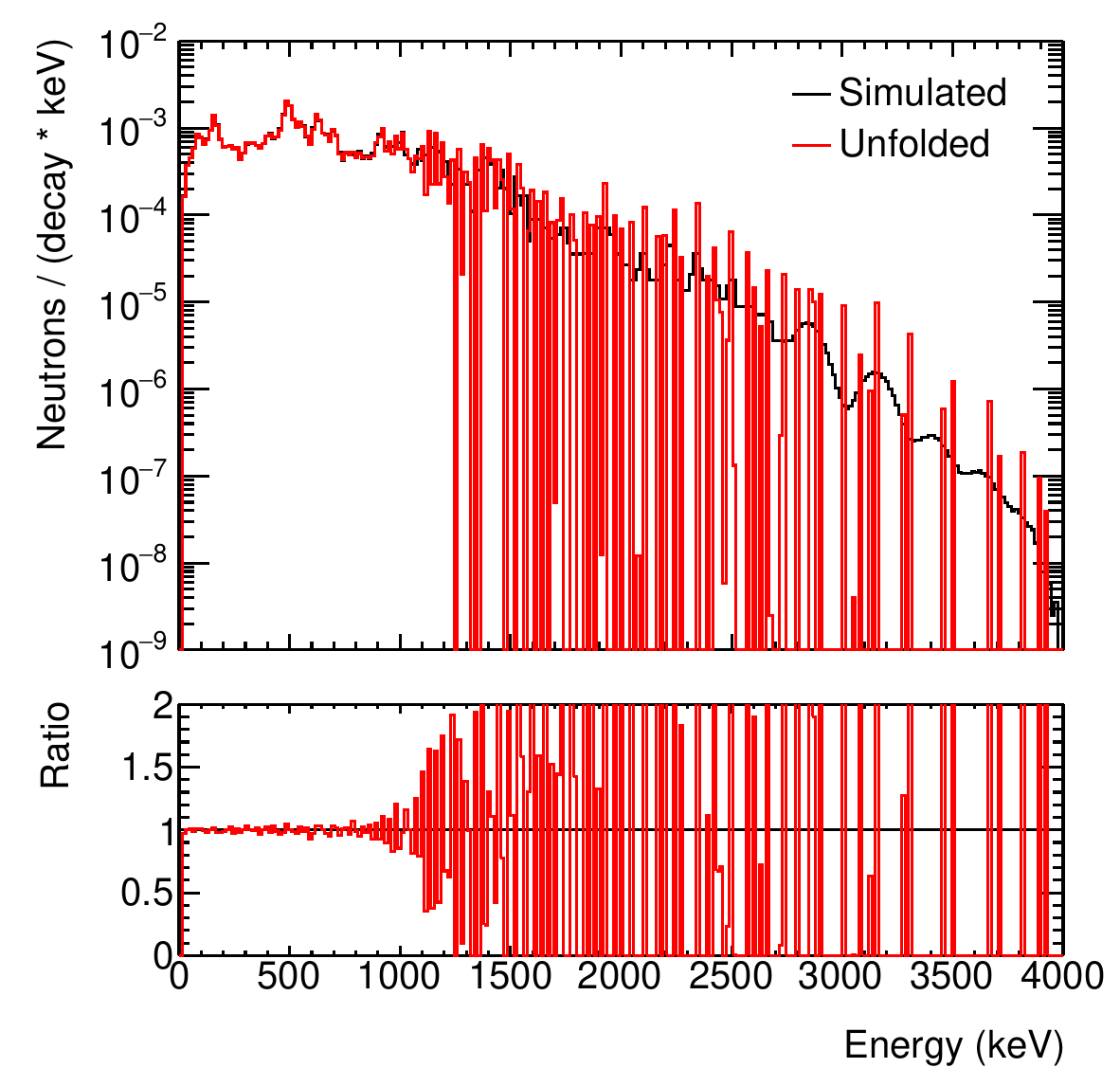}}
  \subfigure[\( \Delta t \)-binning]{\includegraphics[width=0.49\linewidth]{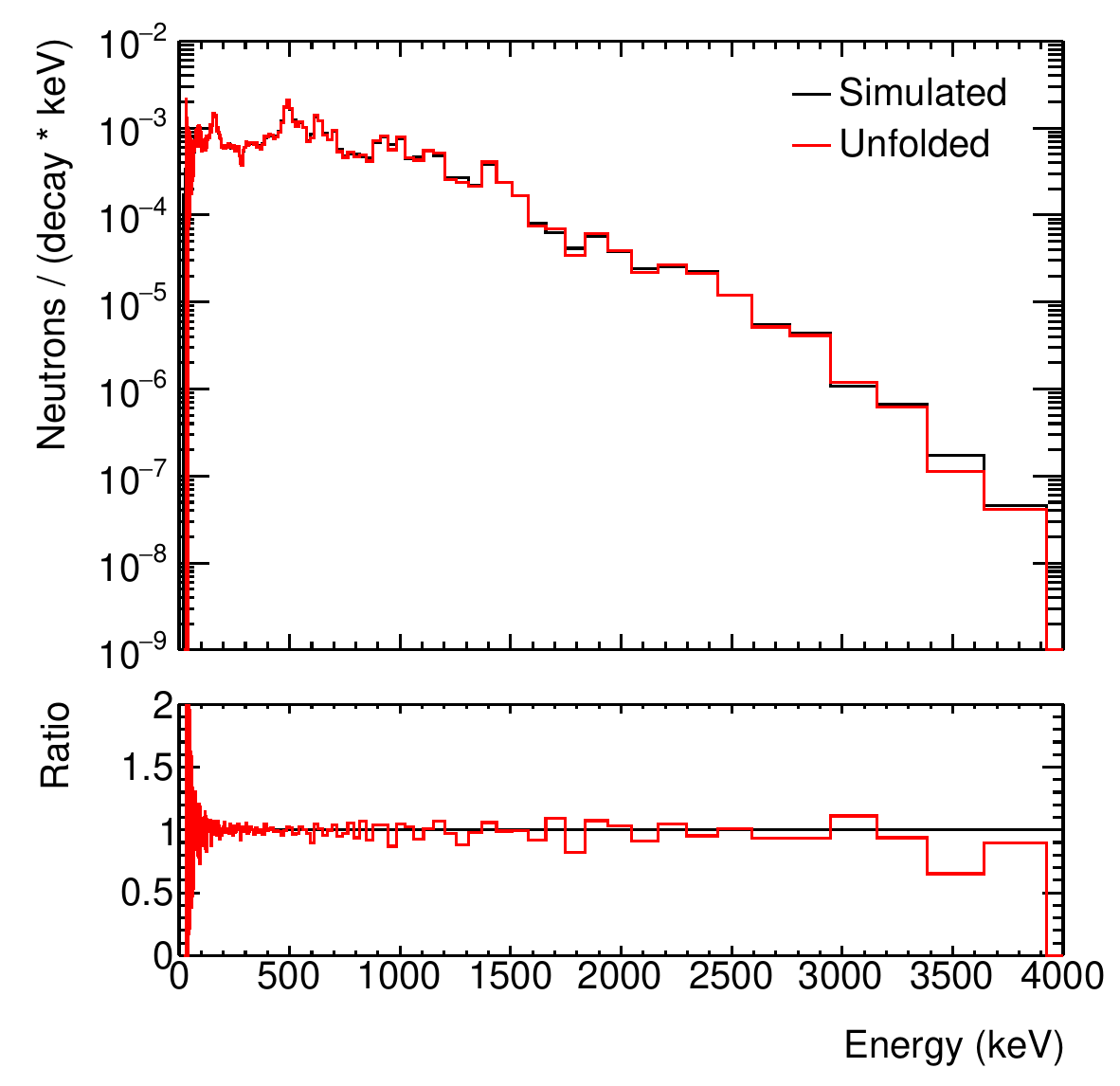}}
  \subfigure[\( \Delta R \)-binning]{\includegraphics[width=0.49\linewidth]{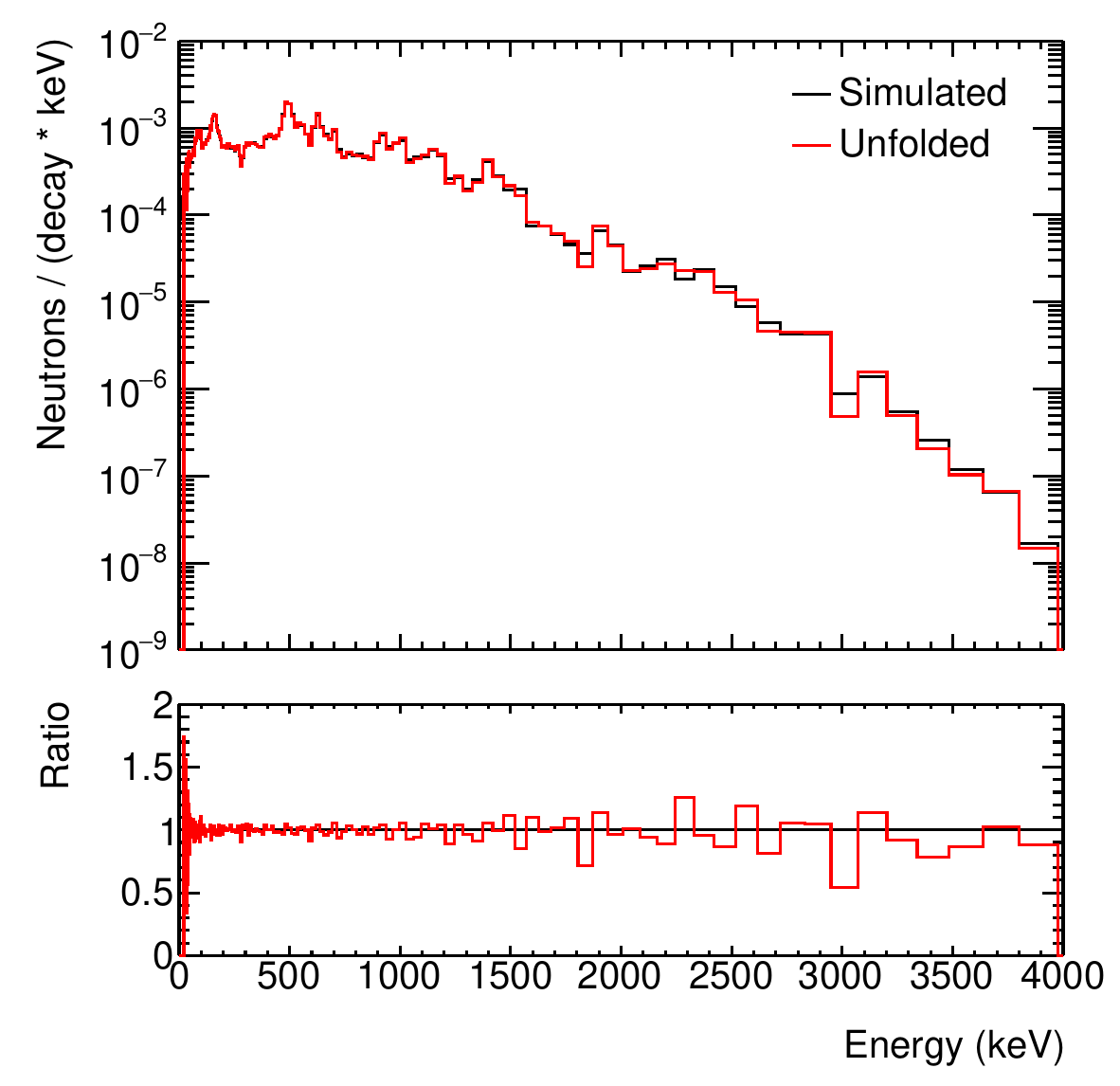}}
  \caption{Neutron energy distributions obtained with the response matrices built for the different types of binning studied (``Unfolded'') compared with the simulated distribution (``Simulated''). In each case, the simulated distribution has been re-binned to the binning of the respective response matrix for easier comparison.}\label{EnergyDistributions}
\end{figure*}

The simulated \( ^{85} \)As TOF spectrum was then unfolded with each of these three response matrices. For simplicity, in this first analysis only the neutron emission part of the \( \beta \)-decay was considered and the effect of the \( \beta \)-detection efficiency was ignored. The reconstructed TOF spectra obtained with each kind of binning can be seen in Figure~\ref{Reconstructions}. They are almost indistinguishable from the simulated ones. In addition to the simulated and reconstructed TOF spectra, the contribution of each individual cause is also shown in the figure. This allows to see that the simulated TOF spectrum can be reconstructed accurately with different sets of responses. Furthermore, this shows that peak-like structures in the TOF spectrum do not necessarily correspond to single transitions, but rather to regions of increased level density or feeding.

Finally, the neutron energy distributions obtained with the three response matrices are compared to the original \( ^{85} \)As neutron energy distribution in Figure~\ref{EnergyDistributions}. The ratios between the unfolded and the original neutron energy distributions are also shown in the figure, for a better visualization of the differences.

In panel (a) of the figure, it can be observed how the \( \Delta E \)-binning offers a good agreement below \( 500 \) keV, but it is not able to reproduce the simulated distribution above \( 900 \) keV accurately. This is because energy differences of \( 15 \) keV at high neutron energies produce too-narrow TOF responses, resulting in an over-sampling of the high-energy region, and thus producing the discretization effect observed.

In contrast, as it can be seen in panel (b) of the figure, the \( \Delta t \)-binning results in a smoother distribution in the entire energy range, but a worse reproduction at low energies. This is because a \( 2.8 \) ns bin width produces too-narrow TOF responses at low neutron energies. It can be observed as well that the high-energy region is described with bins that are too broad, leading to a loss of information. This is because the chosen bin width produces too-broad TOF responses at high neutron energies.

Lastly, the \( \Delta R \)-binning (shown in panel (c) of the figure) offers a better overall reproduction of the original neutron energy distribution over the entire energy range. In this case, the high-energy region is described with narrower bins than in the case of the \( \Delta t \)-binning. Compared to the \( \Delta E \)-binning, the \( \Delta R \)-binning offers a more accurate reproduction of the energy region over \( 900 \) keV. A similar accuracy is obtained in the energy region between \( 50 \) and \( 900 \) keV with both binnings. However, the \( \Delta R \)-binning is not able to reproduce the energy region below \( 50 \) keV as precisely as the \( \Delta E \)-binning.

In conclusion, the \( \Delta R \)-binning was adopted for the analysis, since it allows inferring the neutron energy distribution from the TOF spectrum with the best overall accuracy over the whole neutron energy range, with a ratio very close to \( 1 \).

\subsection{Impact of the neutron detection threshold}\label{thres}

\begin{figure}[tbp]
  \centering
  \subfigure[Reconstruction]{\includegraphics[width=\linewidth]{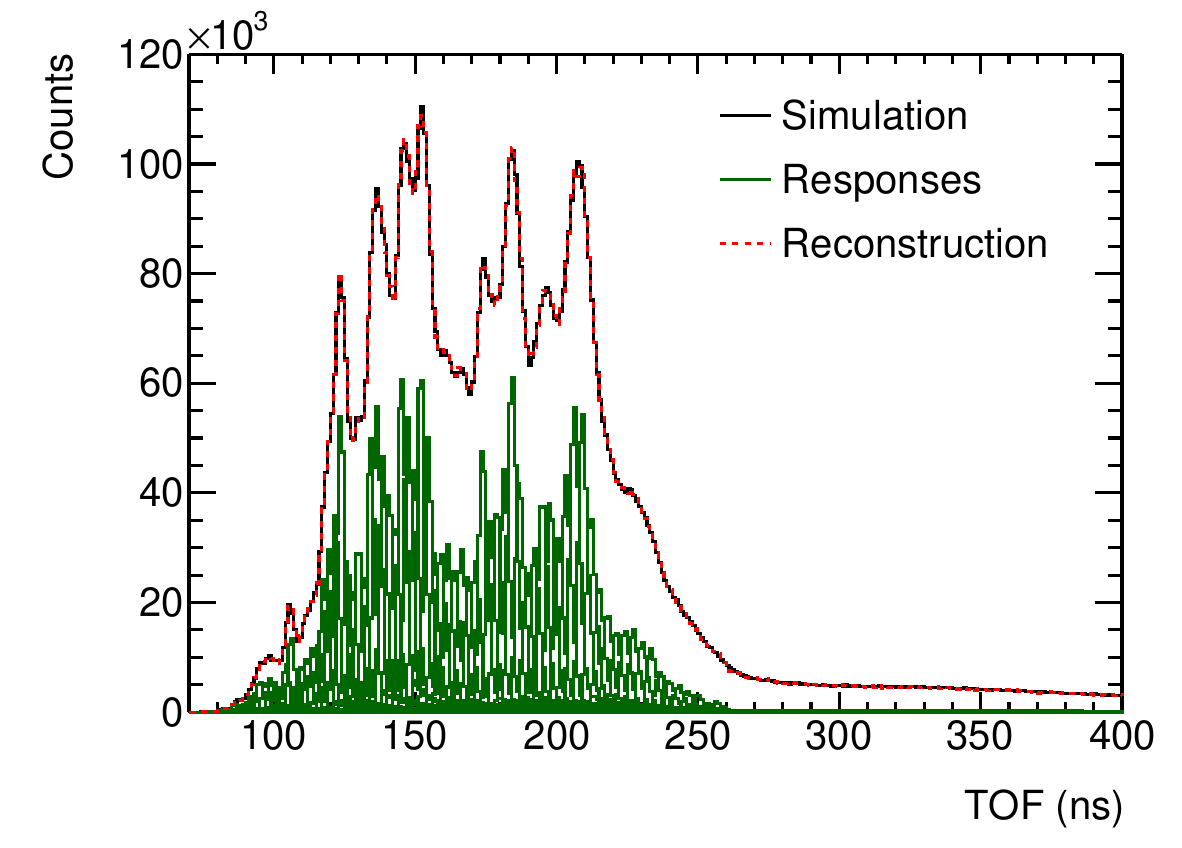}}
  \subfigure[Energy distribution]{\includegraphics[width=\linewidth]{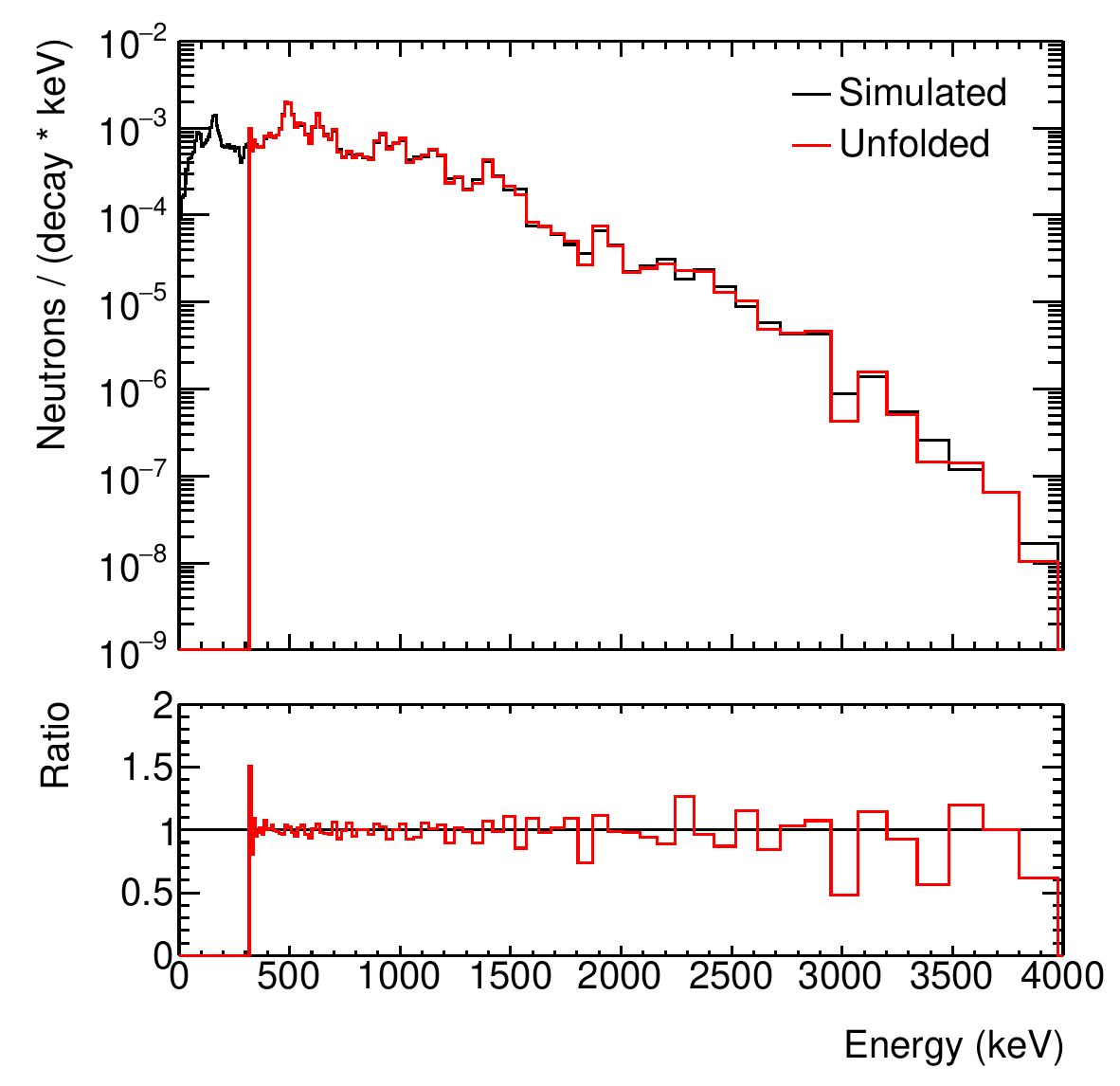}}
  \caption{Unfolding results taking into account a neutron detection threshold of \( \sim 300 \) keV. Panel (a) shows the simulated (``Simulated'', in black) and reconstructed (``Reconstruction'', in dashed red) TOF spectra, as well as the individual responses (``Responses'', in green). Panel (b) shows the neutron energy distribution obtained (``Unfolded'') compared with the simulated distribution (``Simulated'').}\label{Threshold}
\end{figure}

\begin{figure}[tbp]
  \centering
  \subfigure[Reconstruction]{\includegraphics[width=\linewidth]{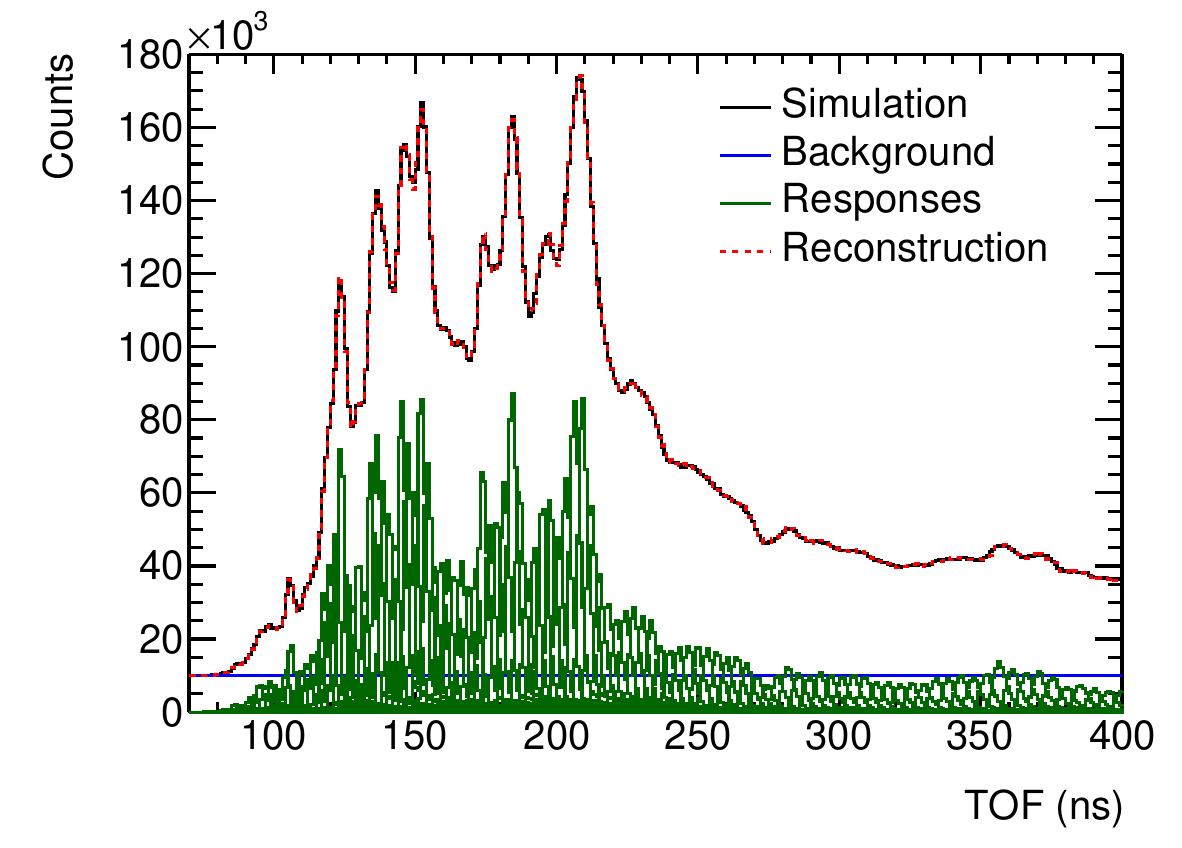}}
  \subfigure[Energy distribution]{\includegraphics[width=\linewidth]{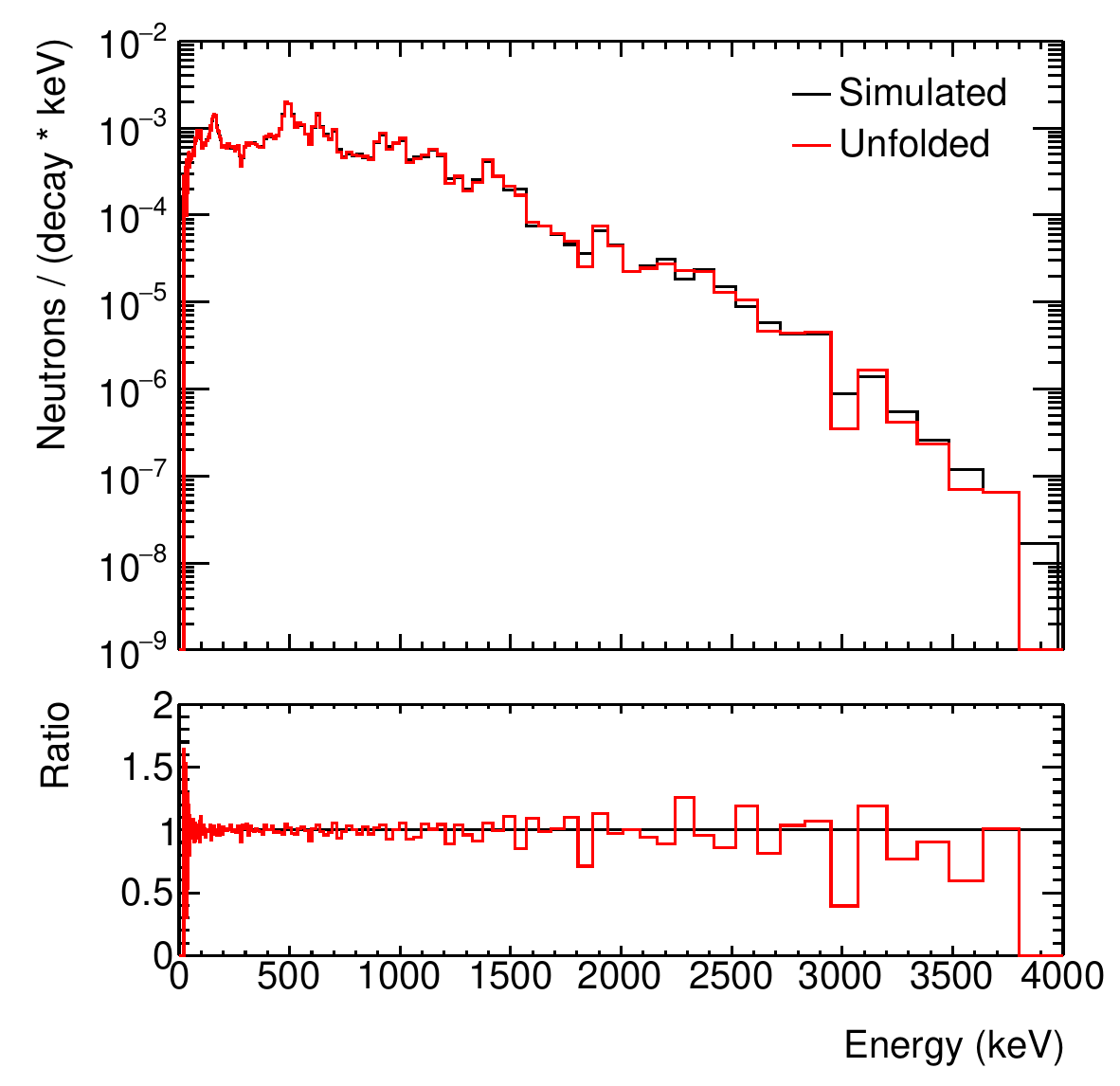}}
  \caption{Unfolding results taking into account an uncorrelated background. Panel (a) shows the simulated (``Simulated'', in black) and reconstructed (``Reconstruction'', in dashed red) TOF spectra, as well as the contribution of the individual responses (``Responses'', in green) and the background (``Background'', in blue). Panel (b) shows the neutron energy distribution obtained (``Unfolded'') compared with the simulated distribution (``Simulated'').\vspace{\baselineskip}}\label{Background}
\end{figure}

\begin{figure}[tbp]
  \centering
  \subfigure[Reconstruction]{\includegraphics[width=\linewidth]{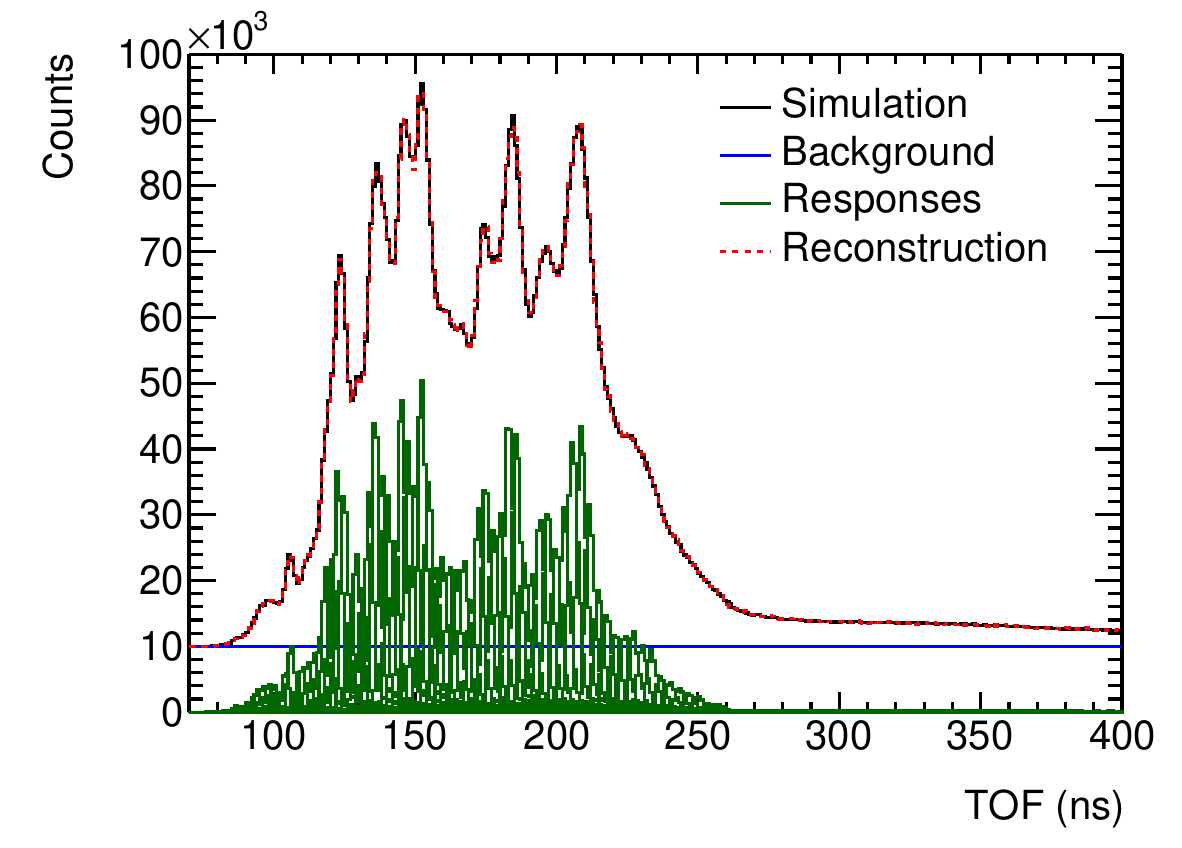}}
  \subfigure[Energy distribution]{\includegraphics[width=\linewidth]{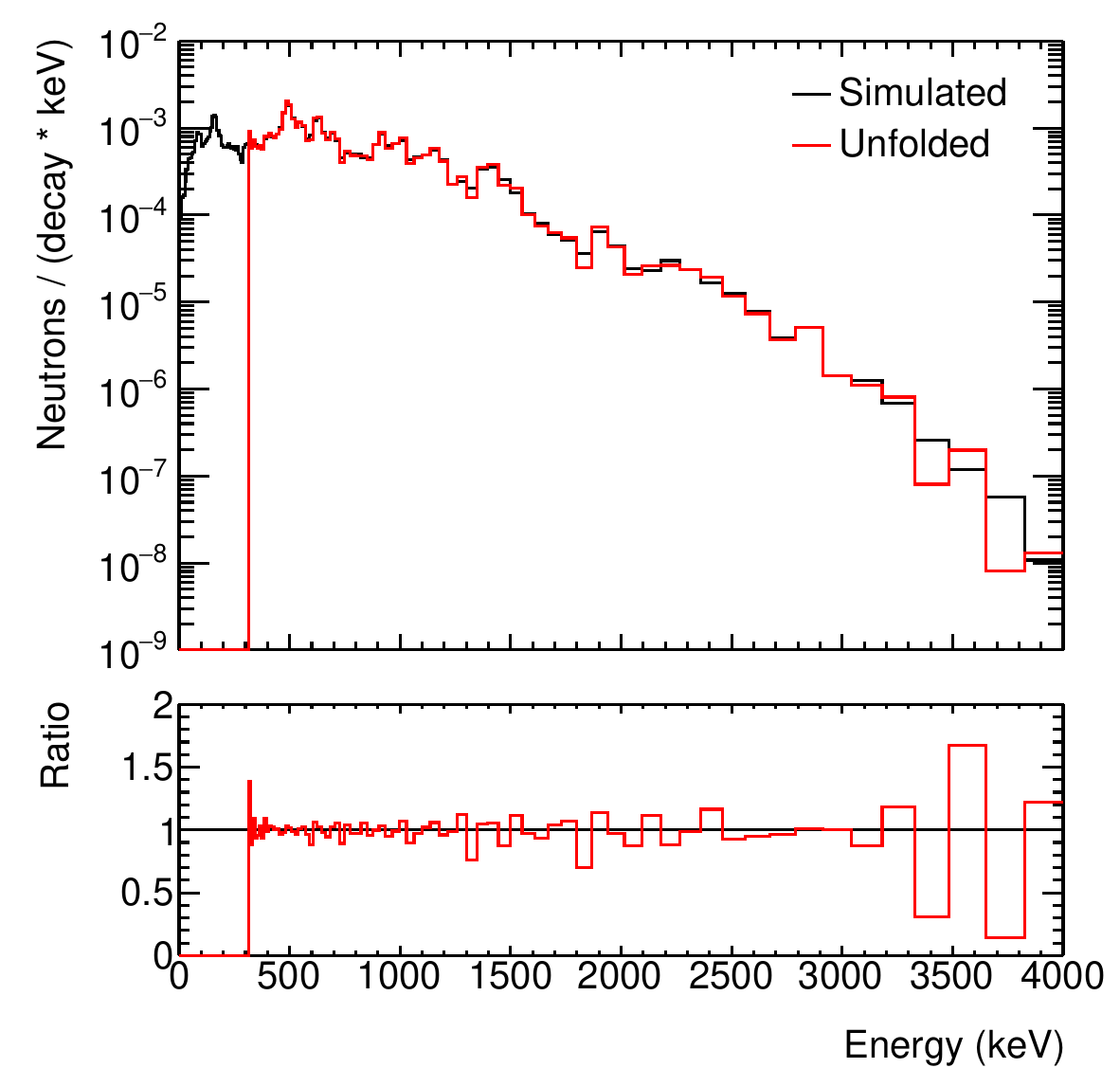}}
  \caption{Unfolding results including the full detection system and taking into account the whole \( \beta \)-decay, the detection thresholds, and an uncorrelated background. Panel (a) shows the simulated (``Simulated'', in black) and reconstructed (``Reconstruction'', in dashed red) TOF spectra, as well as the contribution of the individual responses (``Responses'', in green) and the background (``Background'', in blue). Panel (b) shows the neutron energy distribution obtained (``Unfolded'') compared with the simulated distribution (``Simulated'').}\label{FullThresholdBackground}
\end{figure}

The effect of the neutron detection threshold, as in the case of a real experiment, has also been studied by analyzing the virtual experiment. The same threshold was applied to the simulated neutron detector pulse height spectrum for producing the TOF spectrum and to the response matrix used in the unfolding. In this process, a realistic non-proportional light output of the liquid scintillators was used in the simulations. The effect of the neutron detection threshold is two-fold: it reduces the number of causes producing detectable effects, and the causes that produce them do it over a smaller TOF range. Figure~\ref{Threshold} shows the simulated and reconstructed TOF spectra (top panel) and the resulting neutron energy distribution (bottom panel) after applying a detection threshold of \( 30 \) keVee, corresponding to a neutron energy of \( \sim 300 \) keV. The value corresponds to the real threshold of MONSTER in the experiment~\cite{Thesis}. As can be seen in the figure, the inclusion of the threshold limits the lower neutron energy that can be detected, and hence the original neutron spectrum is never going to be reproduced below that limit. The reconstructed energy distribution has been normalized to the estimated true total number of events \( \hat{N}_{true} \), as calculated from Equation~\ref{TrueEvents}, and the total number of decays \( N_{obs} \).

\subsection{Impact of the background in the TOF spectrum}\label{bckg}

The iterative Bayesian unfolding method allows accounting for the background in a simple way. Spurious correlations that contribute to the TOF spectrum as a flat background can be modeled with an extra cause bin in the response matrix corresponding to a flat time response and with a detection efficiency of \( 1 \). If known, more complex background time structures can also be included in the same manner. The effect of the background in the analysis was investigated by adding a flat background to the simulated TOF spectrum and the mentioned extra cause in the response matrix used for the unfolding.

Figure~\ref{Background} shows in the top panel the simulated and reconstructed TOF spectra, including the background, and in the bottom panel, the energy distribution resulting from the unfolding, compared to the simulated one. In the figure, it can be seen that the inclusion of such a background introduces a sensitivity. It barely affects the neutron energy distribution obtained for favorable signal-to-background ratios, and only starts to deviate from the true distribution at high energies, where neutrons are emitted with low intensity.

\subsection{Analysis of a complete virtual experiment}\label{exp}

To conclude this study, an analysis of a virtual experiment including all the effects previously considered (reference binning, neutron detector threshold, and TOF background) was performed. In this case, the full \( \beta \)-decay was simulated, and the threshold and time resolution of the \( \beta \)-detector were also taken into account. Contrarily to the effect of the threshold in the neutron detectors, the threshold in the \( \beta \)-detector leads to a limited or null detection efficiency of high energy neutrons emitted from unbound states close to the \( Q_{\beta n} \) value of the decay. This is because the \( \beta \)-feeding to high excitation energies corresponds to low \( \beta \)-endpoint energies, close to or below the threshold. This introduces a dependence of the correlated \( \beta\textrm{-}n \) detection efficiency at high neutron energies, which has to be implemented in the analysis.

As can be seen in Figure~\ref{FullThresholdBackground}, the original and reconstructed neutron energy distributions are in excellent agreement over a large energy range, from around \( 300 \) keV onward. The reconstructed energy distribution has been normalized to the estimated true total number of events \( \hat{N}_{true} \) and the detected number of decays \( N_{obs} \). However, it should be said that the introduction of the \( \beta \)-detection efficiency impedes knowing the correct total number of decays. Hence, in a situation with a large \( \beta \)-feeding to high excitation energies, the total areas of the two distributions over the neutron detection threshold would differ, and high-energy neutrons would have not been observed. It can also be seen on the figure that the detection threshold of MONSTER sets a limit in the neutron energies that can be detected and thus can lead to an underestimation of the total number of neutrons emitted in the \( \beta \)-decay. This underestimation will depend on the emission spectrum of the \( \beta \)-decay being measured.

\section{Conclusions}\label{conc}

In this paper, an innovative methodology for obtaining the neutron energy distribution from a TOF measurement based on accurate Monte Carlo simulations and the iterative Bayesian unfolding method has been presented. This methodology presents several advantages over the state-of-the-art methods of obtaining the neutron energy spectrum from a TOF measurement. The method proposed leads to a solution compatible with the data in a statistical sense, avoids numerical problems related to the fit of numerous individual response functions to the data, and can incorporate various systematic effects naturally. The presented methodology has been validated through the analysis of a virtual \( \beta \)-decay experiment:

\begin{itemize}
  \item The dependence of the solution with the binning of the response matrix has been studied. A binning according to the energy resolution of the detection system has been found to yield the most accurate results.
  \item The impact of detection thresholds in the detection system has been evaluated. It has been found that different kinds of thresholds impose different limitations on the energy range that can be reproduced.
  \item The presence of an uncorrelated background has also been studied. This kind of background can limit the ability to detect neutrons emitted with low probability.
  \item A realistic experiment with all the effects combined, including the effect of the \( \beta \)-detector, has been successfully analyzed, yielding accurate results over a wide energy range.
\end{itemize}

This methodology also permits the study of systematic uncertainties by analyzing the data with different response matrices built with variations of the relevant parameters. The results obtained in this way allow quantifying the systematic uncertainties in the result due to the uncertainties in the parameters varied.

The methodology discussed in this paper has been applied successfully to obtain the \( \beta \)-delayed neutron energy spectra emitted in the decays of \( ^{85, 86} \)As. These measurements served as an experimental validation of the methodology and for producing new physics results. These results will be published in a forthcoming paper.

\section*{CRediT authorship contribution statement}\label{cred}

\textbf{A.~Pérez de Rada Fiol:} Writing - Original draft, Methodology, Investigation, Formal analysis, Software. 
\textbf{D.~Cano-Ott:} Conceptualization, Methodology, Supervision, Writing - Review, Project administration, Funding acquisition. 
\textbf{T.~Martínez:} Writing - Review, Investigation, Supervision. 
\textbf{V.~Alcayne:} Investigation.
\textbf{E.~Mendoza:} Writing - Review, Investigation. 
\textbf{J.~Plaza:} Writing - Review, Investigation. 
\textbf{A.~Sanchez-Caballero:} Investigation.
\textbf{D.~Villamarín:} Investigation.

\section*{Declaration of competing interest}\label{decl}

The authors declare that they have no known competing financial interests or personal relationships that could have appeared to influence the work reported in this paper.

\section*{Acknowledgments}\label{ack}

This work was partially supported by the I+D+i grants FPA2016-76765-P, PGC2018-096717-B-C21, and PID2022-142589OB-I00 funded by MCIN/AEI/10.13039/501100011033, and by the European Commission H2020 Framework Programme project SANDA (Grant agreement ID:\ 847552). 

\section*{Data availability}\label{data}

Data will be made available on request.

\bibliographystyle{elsarticle-num}
\bibliography{bibliography}

\end{document}